\documentclass[prl,twocolumn,showpacs,epsfig,aps]{revtex4}
\usepackage{amsfonts}
\usepackage{amsmath}
\usepackage{amssymb}
\usepackage{graphicx}

\providecommand{\U}[1]{\protect\rule{.1in}{.1in}}

\begin{document}

\title{A rate-distortion scenario for the emergence and evolution of noisy
molecular codes }
\author{Tsvi Tlusty }
\affiliation{Department of Physics of Complex Systems, Weizmann Institute of Science,
IL-76100 Rehovot, Israel}

\begin{abstract}
We discuss, in terms of rate-distortion theory, the fitness of molecular
codes as the problem of designing an optimal information channel. The
fitness is governed by an interplay between the cost and quality of the
channel, which induces smoothness in the code. By incorporating this code
fitness into population dynamics models, we suggest that the emergence and
evolution of molecular codes may be explained by simple channel design
considerations.
\end{abstract}

\date{\today }
\maketitle

Living systems store information in one form of a molecule (e.g. DNA) and
use it to produce a different molecule (e.g. protein), usually relying on
intermediary recognition processes by other molecules (e.g. tRNA). This
information transfer is a code, albeit one that must perform in a noisy
environment. Such noisy information channels are omnipresent in biology and
analyzing them can highlight the engineering constraints on living systems
and their impact on fitness. Several studies have examined the biophysical
makeup of the transcription regulatory network (TRN) to scrutinize the
effect of this information system on fitness \cite{SenguptaShraiman02,
GerlandHwa02, BergLassig04}.

In this work, we first introduce a measure for the fitness of molecular
codes. The quality of the code is measured by the distortion of a typical
message. The cost is the typical number of bits required to write one
message. The overall fitness of the code is the weighed sum of cost and
quality. This is similar to the basic problem of rate-distortion theory \cite%
{Shannon59,Berger71} -- how to design an optimal information channel
by balancing the cost against the required transmission quality. We
find that the relevant control parameter is the derivative of cost
with respect to quality, termed gain. The code appears at a phase
transition in the
information channel \cite%
{Berger71,Rose90,Luttrell94,Graepel97,Hofmann97,Rose98,Tishby99}
with the gain playing a role of an inverse temperature. To examine
the appearance of codes we then turn to models describing
populations of information-processing systems, simplified
``organisms", which compete and evolve according to the fitness of
their codes. We show that the coding transition can be induced by
changing a number of parameters such as the accuracy of reading and
the population size. Finally, we treat two realistic scenarios of
deviations from the simplified ideal dynamics, which involve
mutations and genetic drift (i.e. reproduction fluctuations).
Mutations broaden the population to include systems with lower code
fitness, creating a ``quasi-species"\ with a reduced effective
fitness. Genetic drift delays the coding transition to higher gains.

\begin{figure}[th]
\begin{center}
\includegraphics[width=7.6cm]{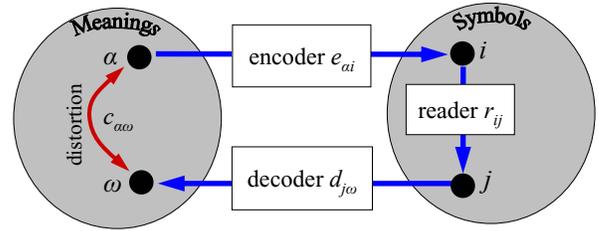} \vspace{0.0cm}
\end{center}
\caption{\textbf{Molecular codes as noisy information channels. }The
code relates the space of meanings (left) with that of symbols
(right). The
channel is a three-stage Markov process (blue arrows): (i) A meaning $%
\protect\alpha $ is encoded as a symbol $i$ by the encoder $e_{\protect%
\alpha i}$ (ii) $i$ is read as $j$ by the reader $r_{ij}$(iii)$j$ is decoded
as $\protect\omega $ by the decoder $d_{j\protect\omega }$. The distance
between the original and the reconstructed meanings is $c_{\protect\alpha
\protect\omega }$ (red arrow). }
\label{Channel}
\end{figure}

\emph{The fitness of molecular codes. -- }Molecular codes often relate two
sets of molecules, which we may think of as \textit{symbols} and their
potential \textit{meanings. }In the genetic code, for example, the symbols
are the 64 DNA\ base-triplets (codons) and their meanings are the 20
amino-acids and the stop signal \cite{Tlusty07JTB}. In the case of the TRN,
DNA\ sites are the symbols and the meanings are the transcription factors
that bind the sites \cite{Shinar06,Itzkovitz06}. Optimizing the quality and
cost of a biological code can therefore be regarded as a semantic problem of
wisely assigning meanings to symbols. To discuss this semantic problem, we
consider an information channel that relates two spaces, one with $s$
symbols and the other with $m$ meanings (Fig. \ref{Channel}). The channel
describes how meanings are stored in memory as molecular symbols, and how
the symbols are read to reconstruct the meaning.

Molecular codes rely on error-prone binding and the channel is therefore
described by a three-stage stochastic process \cite%
{Berger71,Luttrell94,Graepel97,Rose98}:\ (\textbf{i}) The storage of
meanings in memory as symbols is represented by an encoder matrix
$e_{ai}$, the probability that a meaning $\alpha $ is encoded by a
symbol $i$\textit{.} (\textbf{ii}) The symbol is read as described
by the reader matrix $r_{ij}$, the probability to read the symbol
$i$ as $j$, which accounts for possible misreading errors.
(\textbf{iii}) Finally, the read symbol $j$ is
interpreted as carrying a meaning $\omega $ according to a decoder matrix $%
d_{j\omega }$. The distortion between the original meaning $\alpha $ and the
reconstructed one $\omega $ is measured by the distance $c_{a\omega }$. In
the genetic code, for example, amino-acid meanings are encoded as
base-triplet symbols, which in turn are read by tRNAs at the ribosome.
Finally, the decoded amino-acid carried by the tRNA is ligated to the
synthesized protein.

\begin{figure}[tbp]
\begin{center}
\includegraphics[width=7.5cm]{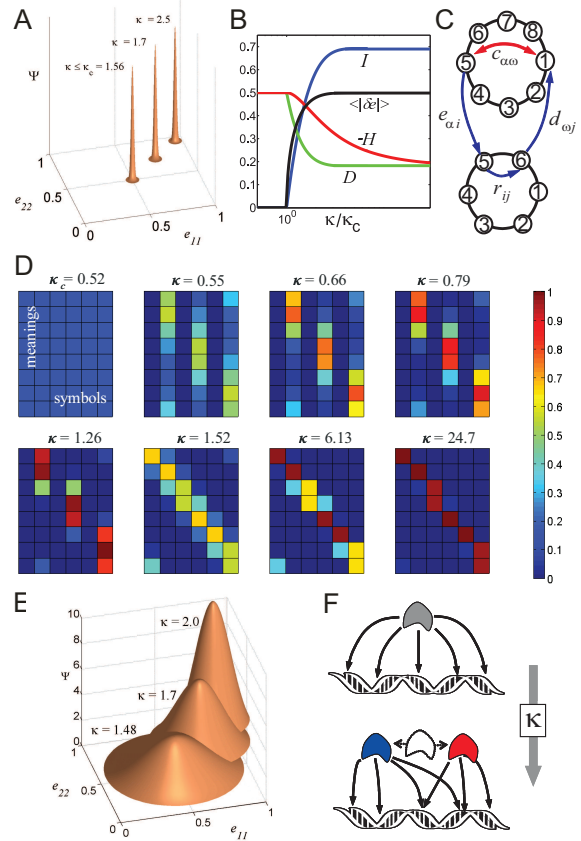} \vspace{0.0cm}
\end{center}
\caption{\textbf{Emergence and evolution of molecular codes. } (\textbf{A})
A code relates $m=2$ meanings $\{1,2\}$ and $s=2$ symbols $\{1,2\}$. The
encoder $e_{\protect\alpha i}$ has $4$ entries and is constrained to a 2D
square by the $2$ conservation relations $e_{11}+e_{12}=e_{21}+e_{22}=1$.
The reader is $r_{ij}=(1-2\protect\varepsilon )\protect\delta _{ij}+\protect%
\varepsilon $ with the misreading probability $\protect\varepsilon =0.1$. At
low mutation rates the population peaks at the optimal encoder (illustrated
as sharp peaks). Below the critical gain $\protect\kappa _{c}=1.56$ the
state is non-coding with $e_{\protect\alpha i}=\frac{1}{2}$. Above $\protect%
\kappa _{c}$ a coding state evolves. (\textbf{B}) The channel cost $I$
increases from $0$ at the coding transition, $\protect\kappa =\protect\kappa %
_{c}$, while the the distortion $D$ decreases. The average order-parameter $%
<|\protect\delta e|>$ increases continuously from $0$ at the second-order
coding transition. The fitness $H$ (plotted is $-H$) increases to an
asymptotic value. (\textbf{C}) A code that relates $8$ meanings and $6$
symbols. The distance is $c_{\protect\alpha \protect\omega }=\min (|\protect%
\alpha -\protect\omega |,8-|\protect\alpha -\protect\omega |)$ and the
reader is defined by a probability of $0.98$ that $i$ is read as $i$ and $%
0.01$ that it is read as one of its two neighbors on the symbol graph. (%
\textbf{D}) The optimal encoder $e_{\protect\alpha i}$ is plotted as
color-coded $6\times 8$ arrays at increasing gains $k$. Below $\protect%
\kappa _{c}=0.52$ (top left) the encoder is $e_{\protect\alpha i}=1/6$ with
uncorrelated symbols and meanings. A\textit{\ coding} state emerges at $%
\protect\kappa _{c}$. The symbol-meaning correlation increases with $\protect%
\kappa $ until every meaning $\protect\alpha $ is encoded by exactly one
symbol $i$ (bottom right). The optimal code is smooth, i.e. close meanings
are encoded by close symbols, as manifested by the continuous diagonal shape
of the encoder. (\textbf{E}) Quasi-species dynamics of the code from \textbf{%
A} with mutation rate $\protect\mu =5\cdot 10^{-5}$. Below $\protect\kappa %
_{c}=1.56$, the population distribution $\Psi $ is smeared around the
non-coding optimum. Above $\protect\kappa _{c}$, a coding state appears, $%
\Psi $ sharpens and migrates towards the one-to-one code $e_{11}=e_{22}=1$. (%
\textbf{F}) A coding transition in the TRN, when a universal
transcription factor splits into distinct species when the gain
$\protect\kappa $ increases. } \label{phase_transition}
\end{figure}

To estimate the \textit{quality} of the coding system one examines
how well a reconstructed meaning preserves the original one. This is
measured by the average \textit{distortion }$D$
\cite{Shannon59,Berger71} along all possible paths $\alpha
-i-j-\omega $ between original and reconstructed meanings
\cite{Luttrell94,Graepel97,Rose98}. The \textit{demand} for each
meaning $\alpha$ is $f_{\alpha }$, which accounts for the
possibility that some meanings are used more frequently than others.
To calculate $D$, each path is weighed by its probability,
$f_{\alpha }e_{\alpha i}r_{ij}d_{j\omega }$, and the summation
yields:
\begin{equation}
D=\left\langle c_{a\omega }\right\rangle =\sum\limits_{\alpha ,i,j,\omega
}f_{\alpha }e_{\alpha i}r_{ij}d_{j\omega }c_{\alpha \omega }.  \label{D}
\end{equation}%
The reader $r_{ij}$ may be represented as a graph, in which the
nodes are the symbols and edges connect symbols that are likely to
be confused (e.g. Fig. \ref{phase_transition}C)
\cite{Tlusty07JTB,Tlusty07Math}. If the reader was ideal
($r_{ij}=\delta _{ij}$) then it would have been advantageous to
decode as many meanings as there are available symbols. However,
since the molecular reader is not perfect it is preferable to decode
fewer meanings and thereby minimize the effect of misreading errors.
Moreover, the preferable codes are \textit{smooth}, that is symbols
that are likely to be confused encode the same meaning or meanings
that are close
with respect to the distance $c_{\alpha \omega }$ \cite%
{Tlusty07JTB,Itzkovitz06,Shinar06,Tlusty07Math} (Fig. \ref{phase_transition}%
D).

A common measure for the \textit{cost} of a coding system is the mutual
information $I$ that estimates the average number of bits required to encode
one meaning \cite{Berger71},%
\begin{equation}
I=\sum_{\alpha ,i}f_{\alpha }e_{\alpha i}\ln \frac{e_{\alpha i}}{u_{i}},
\label{I}
\end{equation}%
where $u_{i}=\sum_{\alpha }f_{\alpha }e_{\alpha i}$ is the overall
probability to use the symbol $i$. In molecular codes $I$ is directly linked
to fitness: The decoder $e_{\alpha i}$ is the probability that the molecule
carrying the meaning $\alpha $ binds the molecular symbol $i$. In the TRN,
for example, $\alpha $ is a transcription factor and $i$ is a prospective
DNA\ binding site. The binding probability $e_{\alpha i}$ scales like a
Boltzmann exponent $e_{\alpha i}\sim \exp \varepsilon _{\alpha i}$, where
the binding energy $\varepsilon _{\alpha i}$ is in $k_{B}T$ units. It
follows that $I$ is actually the average binding energy $I=\sum_{\alpha
,i}f_{\alpha }e_{\alpha i}(\varepsilon _{\alpha i}-\bar{\varepsilon}%
_{i})=\left\langle \varepsilon _{\alpha i}-\bar{\varepsilon}%
_{i}\right\rangle $, with the reference energies $\bar{\varepsilon}_{i}=\ln
\sum_{\beta }f_{\beta }\exp \varepsilon _{\beta i}$. In several molecular
codes (e.g. TRN) the binding energies $\varepsilon _{\alpha i}$ are
approximately linear in the size of the binding sites. The cost $I$ is
therefore proportional to the average size of the binding site \cite%
{SenguptaShraiman02,GerlandHwa02,BergLassig04}. The evolutionary cost to
replicate, transcribe and translate the gene that encodes the binding site
and the cost to correct mutations in this gene, are all expected to be
linear in the binding site size \cite{Alon07}. $I$ is therefore proportional
to the actual fitness cost.

To optimize the molecular coding apparatus, its cost and distortion
must be balanced. We describe this interplay as the maximization of
an overall \textit{code fitness}, $ H=-D-\kappa ^{-1}I$. While $I$
and $D$ are to be minimized, the fitness $H$ is driven by evolution
towards maxima, as manifested by the minus signs in $H$. The
\textit{gain} $\kappa =\partial I/\partial D$ measures the bits of
information required to increase the quality. The gain $\kappa $ is
expected to increase with the complexity the organism and its
environment: The circuitry of a complex organism transmits more
signals and reads a larger genome. It is therefore beneficial for
this organism to pay a larger cost to improve the quality of its
code, since it gains more from such an improvement. Similarly, the
gain is larger in a richer environment.

\emph{Population dynamics in the code space. -- }To examine how
codes evolve in response to changes in the gain, we consider a
population of simplified ``organisms" that compete according to the
fitness of their codes. We imagine a scenario where -- for a given
demand $f_{\alpha }$, determined by the environment, and a given
reader $r_{ij}$ -- each ``organism" has a code specified by its encoder $%
e_{\alpha i}$ and decoder $d_{j\omega }$. The optimal encoder and decoder
are related through Bayes' theorem \cite%
{Luttrell94,Hofmann97,Rose98,Graepel97}, $d_{j\omega
}\sum\nolimits_{\beta ,i}f_{\beta }e_{\beta i}r_{ij}=f_{\omega
}\sum\nolimits_{i}e_{\omega i}r_{ij}$, which states the intuitive
notion that if an encoded meaning $\omega $ tends to be read as the
symbol $j$ then it is likely that $j$ is decoded as $\omega $
\cite{Supp}. Therefore, it suffices to identify every organism by
its encoder $e_{\alpha i}$ and one may describe the population as
points in a ``\textit{code space}", which is spanned by all possible
encoders (Fig. \ref{phase_transition}A). This space is an $m\times
s-$dimensional unit cube $0\leq e_{\alpha i}\leq 1$ and each axis
corresponds to an entry of the encoder $e_{\alpha i}$. An organism
is represented by a point in the cube and the population is a
``cloud" of such points of probability density $\Psi (e_{\alpha
i})$. Since the encoder obeys the $m$ conservation relations
$\sum_{i}e_{\alpha i}=1$, the effective dimension is reduced to
$m\times (s-1)$. In the following, we treat three limiting cases of
population dynamics in the code space: first, a large population
with negligible mutation-rate, next, a large population with
significant mutation rate and, finally, a smaller population with
considerable genetic drift.\

\emph{The coding transition. -- }To find the coding transition we first look
at simplified case of large populations with negligible mutation rate. These
tend to peak at an optimal value of the encoder $e_{\alpha i}^{\ast }$ and
therefore may be approximated by a delta-function, $\Psi (e_{\alpha
i})=\delta (e_{\alpha i}-e_{\alpha i}^{\ast })$. As a result, the dynamics
in this regime amounts to tracing the evolution of the optimal code as the
gain $\kappa $ changes. The optimal code is found at the extremum, $\partial
H/\partial e_{\alpha i}=0$ \cite{Supp}, which leads to
\begin{equation}
e_{\alpha i}^{\ast }=u_{i}e^{-\kappa \Omega _{\alpha
i}}/\sum\nolimits_{j}u_{j}e^{-\kappa \Omega _{\alpha j}}.  \label{optimum}
\end{equation}%
In this Boltzmann partition, the effective energies are $\Omega _{\alpha
i}=\sum\nolimits_{j,\omega }r_{ij}d_{j\omega }(2c_{\alpha \omega
}-\sum\nolimits_{\gamma }d_{j\gamma }c_{\gamma \omega })$ and the gain $%
\kappa $ plays the role of an inverse temperature, i.e. organisms
with lower $\kappa $ are ``hotter" and their codes are noisier.

A simple example for the evolution of a code with increasing gain $\kappa $
is graphed in Fig. \ref{phase_transition}A-B. At low $\kappa $, the optimal
encoder is $e_{\alpha i}=u_{i}$ and $I$ vanishes since the encoder is $%
\alpha -$independent and therefore conveys no information about the meanings
(Eq. \ref{I}). For this reason this state is termed \textit{non-coding}. To
pinpoint the transition, we examine the stability of the fitness with
respect to small variations of the encoder $\delta e_{\alpha i}=e_{\alpha
i}-u_{i}$. The variation $\delta e_{\alpha i}$ is the order-parameter that
describes the emergence of a \textit{coding} state, $\delta e_{\alpha i}\neq
0$, with correlated meanings and symbols. We find that the coding/no-coding
transition takes place at a \textit{critical gain} $\kappa _{c}$ \cite{Supp}%
,
\begin{equation}
\kappa _{c}^{-1}=2\lambda _{R}^{\ast }\lambda _{C}^{\ast },
\label{kappa_c}
\end{equation}%
where $\lambda _{C}^{\ast }$ is the maximal eigenvalue of the normalized
distance, $C_{\alpha \omega }=\sqrt{f_{\alpha }f_{\omega }}(\sum_{\beta
}f_{\beta }c_{\beta \omega }+\sum_{\gamma }f_{\gamma }c_{\alpha \gamma
}-\sum_{\beta \gamma }f_{\beta }f_{\gamma }c_{\beta \gamma }-c_{\alpha
\omega })$, and $\lambda _{R}^{\ast }$ is the second-largest eigenvalue of
the weighted square of the reader $R_{ij}=\sqrt{u_{i}u_{j}}%
\sum_{k}(r_{ik}r_{kj}/\sum_{t}u_{t}r_{tk})$ \cite{Supp}. $\lambda _{R}^{\ast
}$ corresponds to the \textit{smoothest} non-uniform eigenvector $\delta
e_{\alpha i}^{\ast }\neq 0$, which represents a coding state \cite%
{Tlusty07Math,Tlusty07JTB,TlustyTBP}. This eigenvector, which
emerges at the coding transition (Fig. \ref{phase_transition}B-D),
is the first-excited state of the system and measures the tendency
of a meaning $\alpha $ to be encoded by the symbol $i$. Boltzmann
partitions and consequent phase transitions are common in
rate-distortion theory and analogous optimization problems in the
context of clustering, deterministic annealing
and self-organizing maps \cite%
{Shannon59,Berger71,Rose90,Luttrell94,Graepel97,Hofmann97,Rose98,Tishby99}%
.

The critical gain (Eq. \ref{kappa_c}) indicates three possible pathways from
the random, non-coding state towards the emergence of a code: via increasing
the gain $\kappa $, via increasing the reading accuracy (larger $\lambda
_{R} $) or via increasing the average distance between meanings (larger $%
\lambda _{C}$). We suggest that such simple coding/non-coding transitions
may describe the emergence of biological codes. In the case of TRN, for
example, one imagines the primordial circumstances when a primitive organism
had only one universal transcription factor that binds all DNA sites (Fig. %
\ref{phase_transition}F). Then, as $\kappa$ increases, for example
the environment becomes richer in information, the factor splits
into several distinct factors, each binding to specific sites. In
the case
of the genetic code, a series of transitions (like those in Fig. \ref%
{phase_transition}D) is thought to describe the emergence and
evolution of the code \cite{Tlusty07JTB,Tlusty07Math,TlustyTBP}.

\emph{Effects of mutations.-- }Mutations add another kind of noise, smearing
the population over a larger region of the code space. When the mutation
rate $\mu $ is significant one may model the population in terms of
reaction-diffusion dynamics, in the spirit of the quasi-species model \cite%
{Eigen71},%
\begin{equation}
\frac{\partial \Psi }{\partial t}=\left[ H(e_{\alpha i})-\bar{H}\right] \Psi
+\mu \sum_{\alpha ,i}\frac{\partial ^{2}\Psi }{\partial e_{\alpha i}^{2}}.
\label{quasi}
\end{equation}%
In Eq. \ref{quasi}, each organism in the population reproduces at a rate
equal to the fitness of its code $H(e_{\alpha i})$ (the reaction term).
However, codes may mutate at a rate $\mu $. This random walk in the code
space is described by the diffusion term. The fitness $H$ is normalized by
the average fitness $\bar{H}=\int \Psi (e_{\alpha i})H(e_{\alpha
i})de_{\alpha i}$ to ensure conservation of the probability distribution.
Typically, $\Psi $ approaches a steady-state, which corresponds to the
eigenmode of maximal $\bar{H}$ \cite{Eigen71}. To find the steady-state we
approximate the fitness by a quadratic expansion around an optimum $H\simeq
H_{\ast }-%
%TCIMACRO{\U{bd}}%
%BeginExpansion
{\frac12}%
%EndExpansion
\sum\nolimits_{\alpha ,i,\omega ,j}Q_{\alpha i\omega j}\delta
e_{\alpha i}\delta e_{\omega j}$. Assuming a Gaussian ansatz for
$\Psi $ we find the steady-state \cite{Supp},
$\Psi \sim \exp [ -(8\mu) ^{-1/2}\sum_{\alpha ,i,\omega ,j}%
\sqrt{Q}_{\alpha i\omega j}\delta e_{\alpha i}\delta e_{\omega j}]$,
where $\sqrt{Q}$ is the square root of the Hessian $Q_{\alpha
i\omega j}$. $\Psi$ indicates that the mutations smear the
population over a
width that increases with the mutation-rate as $\sim \mu ^{1/4}$(Fig. \ref%
{phase_transition}E), which may be significant even for relatively low
mutation rates due to the small exponent. The leakage by mutations from the
optimal code to lesser codes reduces the average fitness, $\bar{H}=H_{\ast
}-(\mu /2)^{1/2}$Tr$\sqrt{Q}$ \cite{Supp}. At the coding transition (Eq. \ref%
{kappa_c}), the Gaussian $\Psi $ becomes infinitely wide in the
direction of the emergent coding eigenvector $\delta e_{\alpha
i}^{\ast }$, a precursor of the appearance of a coding state along
this direction.

\emph{Effects of genetic drift.-- }The quasi-species dynamics is
deterministic in the sense that it neglects random reproduction
fluctuations, termed genetic drift, which are irrelevant in large
populations. However, when the effective size of the population $n$
is small, $n\mu \ll 1$ -- considered to be the relevant condition
during the emergence of the genetic code, for example -- genetic
drift is a major determinant. The typical dynamics in this regime
exhibits long periods of time when the population resides in the
vicinity of a fitness optimum separated by short transients of
diffusion by genetic drift to another optimum. For our purpose, it
is convenient to coarse-grain this dynamics in space and time and
regard it as instantaneous random transitions between the optima. In
this type of dynamics the distribution $\Psi $ approaches
asymptotically a Boltzmann partition $\Psi \sim e^{nH}$, with an
``inverse temperature" that is equal to the population size $n,$ up
to a factor of order unity \cite{BergLassig04,Sella05}. It is
convenient to define a free energy \cite{Sella05}, $F=\left\langle
-H\right\rangle -n^{-1}S$, in which the fitness is minus the
Hamiltonian and the entropy of the genetic drift is $S=-\int \Psi
\ln \Psi de_{\alpha i}$. A mean-field treatment yields the
approximation (akin to a mean-field Potts model) \cite{Supp},
\begin{equation}
F=-H(\bar{e}_{\alpha i})+n^{-1}\sum_{\alpha ,i}f_{\alpha }\bar{e}_{\alpha
i}\ln f_{\alpha }\bar{e}_{\alpha i},  \label{mean-field}
\end{equation}%
where $\bar{e}_{\alpha i}$ is the average encoder $\bar{e}_{\alpha i}=\int
e_{\alpha i}\Psi (e_{\alpha i})de_{\alpha i}$. Eq. \ref{mean-field}
indicates that the genetic drift contribution $S$ adds another source of
randomness to that of the cost $I$; both drive the system towards the random
non-coding state.

From stability analysis of $F$ it follows that the genetic drift
shifts the critical transition to higher gains, $\kappa
_{c}^{-1}+n_{c}^{-1}=2\lambda _{R}^{\ast }\lambda _{C}^{\ast }$
\cite{Supp}. This also adds a fourth pathway towards the coding
transition, via population growth, to the three pathways suggested
by Eq. \ref{kappa_c}. To give an order-of-magnitude estimate for
$\kappa _{c}$ and  $n_{c}$, we notice that if the misreading
probability\ is relatively small (the non-diagonal terms $R_{ij}\ll
1$) then $\lambda _{R}^{\ast }\simeq 1$. It follows that the smaller
of $\kappa _{c}$ and $n_{c}$ is of
the order of $1/\lambda _{C}^{\ast }$, which roughly scales like the $1/$%
(fitness reduction by one reading error). Such cost-quality
considerations are generic \cite{Savir07,TlustyTBP} and may help to
understand the evolution of other biological information-processing
systems.

TT thanks A. J. Libchaber, E. Moses and J. -P. Eckmann for valuable
discussions and the support of the Minerva fund and the Center for
Complexity Science.

\end{document}